\documentclass[aps,pre,twocolumn,superscriptaddress,showpacs]{revtex4-1}

\usepackage{graphicx,color}
\usepackage{amsmath}
\usepackage{amssymb}
\usepackage{bm}
\usepackage{bbm}
\usepackage[normalem]{ulem}

\def\be{\begin{equation}}
\def\ee#1{\label{#1}\end{equation}}

\def\a{\alpha}
\def\b{\beta}

\def\G{\Gamma}
\def\d{\delta}
\def\D{\Delta}

\def\l{\lambda}

\def\x{\xi}

\def\r{\rho}

\def\t{\tau}

\def\i{\int}

\def\bx{{\mathbf x}}
\def\bX{{\mathbf X}}

\def\bq{{\mathbf q}}
\def\bp{{\mathbf p}}
\def\bz{{\mathbf z}}
\def\bZ{{\mathbf Z}}
\def\by{{\mathbf y}}
\def\bY{{\mathbf Y}}

\def\cN{{\mathcal N}}

\def\neqq{\text{neq}}
\def\Tr{\mbox{Tr}}
\begin{document}
\title{Open system trajectories  specify fluctuating work but not heat}
\author{ Peter Talkner}
\affiliation{Institut f\"{u}r Physik, Universit\"{a}t Augsburg, Universit\"{a}tsstra{\ss}e 1, D-86159 Augsburg, Germany}
\affiliation{Institute of Physics, University of Silisia, 40007 Katowice, Poland}
\author{Peter H\"anggi}
\affiliation{Institut f\"{u}r Physik, Universit\"{a}t Augsburg, Universit\"{a}tsstra{\ss}e 1, D-86159 Augsburg, Germany}
\affiliation{Nanosystems Initiative Munich, Schellingstr, 4, D-80799 M\"unchen, Germany} 
\affiliation{Department of Applied Mathematics, Lobachevsky State University of Nizhny Novgorod,\\ Nizhny Novgorod 603950,Russia}
\date{\today}

\begin{abstract}
Based on the explicit knowledge of a Hamiltonian of mean force, the classical statistical mechanics and equilibrium thermodynamics of open systems in contact with a thermal environment at arbitrary interaction strength can be formulated. Yet, even though the Hamiltonian of mean force uniquely determines the equilibrium phase space probability density of a strongly coupled open system  the knowledge of this probability density alone is insufficient to determine the Hamiltonian of mean force, needed in constructing the underlying  statistical mechanics and thermodynamics. We demonstrate that under the assumption that the Hamiltonian of mean force is known, an extension of thermodynamic structures from the level of averaged quantities to fluctuating objects (i.e. a stochastic thermodynamics) is possible. However, such a construction undesirably involves also a vast ambiguity. This situation is rooted in the eminent lack of a physical guiding principle allowing to  distinguish a physically meaningful theory  out of a multitude of other equally conceivable ones. 
\end{abstract}

\maketitle

\section{Introduction}
Thermodynamics originally evolved from the challenge of how to understand and optimize steam engines. It soon transcended its engineering origin and developed into an abstract phenomenological theory that does not rely on the specific  properties of the systems to which it is applied. Even more, according to Einstein ``Thermodynamics  is the only physical theory which I am convinced will never be overthrown, within the framework of applicability of its basic concepts'' \cite{Einstein}.

The field of thermodynamics is based on the idea of thermodynamic equilibrium, describing a steady state that is characterized by a very small number of relevant macroscopic variables such as energy, volume, particle numbers and order parameters in the case of broken symmetries \cite{Callen}. Irreversible thermodynamics in turn includes the  description of time-dependent phenomena. It, however, is based on the assumption of local thermal equilibrium  and hence is restricted to processes close to equilibrium; as such it presents a phenomenological, often very useful approach \cite{dGM}.

The question whether thermodynamic principles possibly also rule far-from-equilibrium situations has a long history \cite{KP}, -- although without a generally accepted answer up to date. More recent attempts to impose thermodynamic structures on the  {\it trajectory level of stochastic processes 
} have been put forward under the names of {\it stochastic energetics} \cite{SekimotoPTP,SekimotoLNP} and {\it stochastic thermodynamics} \cite{SeifertAIP,SeifertRPP}.  For the implementation of both approaches an
energy-like quantity needs to be defined as a function on the state-space of the considered system. A Boltzmann-type probability density specified by an ambient inverse temperature $\b$ multiplying this energy expression is supposed to characterize the distribution of states in thermal equilibrium. The equilibrium average of the energy expression is understood as the internal energy of the considered system.

Because in these approaches the state of the system undergoes a stochastic process due to the interaction between  system and environment the energy function becomes a time-dependent fluctuating quantity. It hence is termed a fluctuating internal energy. A central assumption  of stochastic energetics and stochastic thermodynamics  is that the fluctuating internal energy not only characterizes the system in its equilibrium state specified by the above mentioned Boltzmann distribution but also covers a class of nonequilibrium situations.
Within this class the environment consists of a single heat bath at an inverse temperature $\b$. It contains relaxation processes emanating from  a nonequilibrium initial state  of the open system as well as processes that are driven by a time-dependent variation of system parameters $\l$.  
Typically, the resulting dynamics is modeled  with an overdamped Langevin dynamics \cite{SekimotoPTP,SekimotoLNP,SeifertPRL,SeifertAIP,SeifertRPP}.

The deterministic part of the Langevin dynamics may depend on externally controllable  parameters leading upon variation to an energy change, which is then interpreted as work applied to the system. In a first-lawlike fashion the difference between fluctuating internal energy and fluctuating work is considered as fluctuating heat, both in the framework of stochastic energetics and stochastic thermodynamics. Stochastic thermodynamics proceeds one step further and also introduces a fluctuating entropy; the latter is essentially given by the logarithm of the instantaneous probability density of the system \cite{SeifertPRL}. Even though it was noted by Sekimoto \cite{SekimotoR} that the potential landscape in which a stochastic motion takes place is in fact a constrained free energy, its possible temperature dependence has been ignored for both, stochastic energetics and stochastic thermodynamics, a notable exception is the recent  work  in Ref.  \cite{Seifert}.

The statistical mechanics and thermodynamics of open systems that interact at a finite strength with their environment \cite{FV,FLO,ActaPol,HIT,IHT}  cannot be treated within the usual weak coupling framework \cite{weakcoup}. Instead, the equilibrium statistical mechanics is now  governed by a so termed  Hamiltonian of mean force \cite{Kirkwood,HTB,RouxSimonson}. This quantity replaces the bare, microscopic system Hamiltonian  that characterizes the thermodynamics of the considered system staying in  very weak contact to its environment 
with an effective Hamiltonian that
typically depends on both,  the temperature of the environment and additionally also on the coupling-strength between the system and the environment as well as on other properties of the environment. At finite coupling strengths to the heat bath, the internal energy of the open system is given not only by the equilibrium average of the potential of mean force but also contains the average of the temperature derivative of the Hamiltonian of mean force. This particular relation has been transferred in Ref. \cite{Seifert} into the framework of stochastic thermodynamics.

With the present study we investigate the question whether the principles of thermodynamics are sufficient to construct a theory of fluctuating thermodynamic potentials. Our main finding is that there exists a large variety of different families of fluctuating potentials, all being thermodynamically consistent. A physical principle, however, which would allow  to reduce this ambiguity in singling  out a unique physically meaningful theory is lacking.

The paper is organized as follows. In Section \ref{OS} we review the statistical mechanics and thermodynamics of open equilibrium systems and stress that the potential of mean force in general cannot be inferred from the mere knowledge of the reduced probability density of the open system. As an additional information needed to determine the Hamiltonian of mean force  one must know the free energy of the system, being the difference of the free energies of the total system  and of the bare environment. Based on the assumed knowledge of the Hamiltonian of mean force we consider in Section \ref{FIE} the  fluctuating internal energy as it was postulated  recently in  Ref. \cite{Seifert}. That specific choice is not unique, however, as there exist variants thereof, all of which yield the same equilibrium average coinciding with the equilibrium internal energy of the open system. Based on the definition of the fluctuating internal energy the corresponding energy content of the environment can be quantified. The work applied to the system by a change of a system parameter can then be expressed as the difference of the energies of the total system at the beginning and the end of the forcing of the system by means of a parameter change following a prescribed protocol, or, equivalently, by integrating the instantaneous power supplied to the system over the complete time of the forcing \cite{JarzynskiJSM}.
Assuming the validity of a first-lawlike balance relation the difference between the internal energy change and the supplied work is taken as the energy exchanged with the environment and hence interpreted as heat exchange. In Section \ref{Attempt} we review this scenario, as well as an alternative approach in which the energy balance is treated in terms of fluxes. This though leads to an explicitly fluctuating expression for the heat-flux which thus is not accessible on the level of the open system dynamics.

Next, in Section \ref{FEFE} we find that there exists a large manifold of families of thermodynamically consistent fluctuating internal and fluctuating free energies with a matching fluctuating entropy.  Again,  a physically motivated  guiding principle providing  a unique choice is lacking. Therefore, the existence of thermodynamically consistent fluctuating potentials may be looked upon as a curiosity  without possessing a profound physical meaning. Consequently, such  approaches  hardly can be put to physical use uncritically. The study closes with a summary and conclusions,  together with  several appendices on more technical aspects.

With  the present work we restrict ourselves to the consideration of classical systems only; quantum systems pose additional  subtleties and challenges \cite{CHT,THnp,THpre}.

\section{Thermodynamics of  open systems }\label{OS}
An open thermal system can be closed by considering the dynamics of the degrees of freedom of the open system together with all environmental degrees of freedom interacting directly or indirectly with those of the former one.

The total system is then described by a phase space $\G_{\text{tot}} = \G_S \otimes \G_B$ where $\G_S$ and $\G_B$ are the phase spaces of the system and the environment, respectively.
Points in the total phase space are denoted by $\bz =(\bx,\by) \in \G_{\text{tot}}$, where $\bx \in \G_S$ and $\by \in \G_B$ specify the components in the phase spaces of the open system and the environment, respectively.
Accordingly, the dynamics of the total system is governed by a Hamiltonian $H_{\text{tot}}$ that can be written as
\be
H_{\text{tot}}(\bx,\by) = H_S(\bx) + H_i(\bx,\by) + H_B(\by)\:,
\ee{Htot}
where $H_S$, $H_B$ and $H_i$ are the Hamiltonians of the isolated system, the isolated environment and the mutual interaction of the system and environment, respectively.

We assume that the total system stays in thermal equilibrium at the inverse temperature $\b$ and consequently it is described by the canonical probability density function (pdf)
\be
\rho_\b(\bz) = Z^{-1}_{\text{tot}} e^{-\b H_{\text{tot}}(\bz)}\:,
\ee{rb}
where
\be
Z_{\text{tot}}= \int d\G_{\text{tot}}e^{-\b H_{\text{tot}}(\bz)}
\label{Z}
\end{equation}
denotes the partition function of the total system. The infinitesimal phase space volume elements $d\G_S$, $d\G_B$ of the system and the environment, respectively, yielding $d\G_{\text{tot}}=d\G_S d\G_B$  are supposed to be scaled dimensionless. For example in the case of a system consisting of $N$ particles with positions $\bq$ and momenta $\bp$ in a $d$ dimensional configuration space, this is conveniently achieved by the multiplication of the dimensional volume element $d^{dN} \bq d^{dN} \bp $ with the factor $h^{-dN}$ where $h$ is Planck's constant. The volume elements may additionally contain symmetry factors to account properly for the Gibbs paradox in the limit of a large total particle number. The equilibrium state of the open system is  described by the reduced pdf $p_\b(\bx)$  given by
\be
p_\b(\bx) = \int d \G_B \rho_\b(\bx,\by)\:.
\ee{pr}

In order to characterize the thermodynamics of the open system we introduce the notion of the  ``Hamiltonian of mean force'' $H^*$. It   is defined in terms of the average of $e^{-\b (H_S +H_i)}$ with respect to the bare environment \cite{Kirkwood,FV,FLO,HTB,RouxSimonson}, i.e.,
\be
e^{-\b H^*(\bx)} = \langle e^{-\b (H_S(\bx)+H_i(\bx,\by))} \rangle_B\:,
\ee{eH*}
where $\langle \cdot \rangle_B = Z_B^{-1} \int d \Gamma_B \cdot e^{-\b H_B}$ denotes the equilibrium average over the environmental degrees of freedom in the absence of the system. Accordingly,
\be
Z_B = \int d\G_B e^{-\b H_B}
\ee{ZB}
is
the partition function of the bare environment.   For the Hamiltonian of mean force we therefore find
\be
H^*(\bx) = H_S(\bx) - \b^{-1} \ln \langle e^{-\b H_i(\bx,\by)} \rangle_B\:.
\ee{H*}
In general $H^*(\bx)$ manifestly deviates from the bare system Hamiltonian $H_S$. A prominent exception is the case of very weak  coupling between system and environment. Then,  the renormalization of the system Hamiltonian due to environmental degrees of freedom becomes negligible. The classical Zwanzig-Caldeira-Leggett system-bath Hamiltonian already has built in a counter term such that even at strong coupling between system and environment the Hamiltonian of mean force agrees with the bare system Hamiltonian \cite{Zwanzig,CL,Bogolyubov1945}.  This property is shared by the more general class of total Hamiltonians for which the interaction between system and environment can formally be removed by a canonical transformation of the environmental degrees of freedom.

The difference between the Hamiltonian of mean force and the bare system Hamiltonian in general depends on temperature with the exception of harmonic environmental models for which the correction term $-\b^{-1} \ln \langle e^{-\b H_i } \rangle_B$ is always independent of temperature, such as for the bilinear coupling  model to a heat bath of harmonic oscillators  by  Magalinski\u{i} and Ullersma \cite{Magalinskii,Ullersma}. On the other hand, a possible external parameter dependence that enters the total Hamiltonian via $H_S(\bx)$,  is contained in $H^*$ only through $H_S(\bx)$, whereas the temperature dependent part is independent of such a parameter.

The reduced PDF of the open system as defined in Eq. (\ref{pr}) can be expressed in terms of the Hamiltonian of mean force as
\be
p_\b(\bx) = Z^{-1}_S e^{-\b H^*(\bx)}\:,
\ee{pH*}
where
\be
Z_S =\frac{Z_{\text{tot}}}{Z_B} = \int d\Gamma_s e^{-\b H^*}\:.
\ee{ZS}
With this particular form of the partition function $Z_S$ one obtains a consistent thermodynamic description of an open system {\it independent} of the strength of the interaction between system and environment \cite{HIT,IHT}. Applying the standard statistical mechanical rules we obtain for the free energy $F_S$ of the open system
\be
F_S = - \b^{-1} \ln Z_S\:.
\ee{FZ}
Accordingly, also the internal energy $U_S$ and the entropy $S_S$ are defined in their standard thermodynamic form, reading
\begin{align}
\label{US}
U_S&=-\frac{\partial}{\partial  \b} \ln Z_S\:,\\
S_S &= k_B \ln Z_S - k_B \b \frac{\partial}{\partial \b} \ln Z_S\:.
\label{SS}
\end{align}
As a consequence, thermodynamic entropy, free and internal energy are related by
\be
F_S = U_S - T S_S\:,
\ee{FUTS}
further entailing the relations
\begin{align}
U_S &= \frac{\partial}{\partial \b} (\b F_S) \:,
\label{UF}   \\
S_S &= k_B \b^2 \frac{\partial}{\partial \b} F_S\:.
\label{FS}
\end{align}
The validity of any two of the three relations (\ref{FUTS}) --  (\ref{FS}) implies the third one. This feature constitutes the {\it thermodynamic consistency} of the potentials $F_S$, $U_S$ and $S_S$, describing the thermodynamic equilibrium of an open system. This consistency follows from the fact that all these  potentials are implied by the set of equations (\ref{FZ}) -- (\ref{SS}) in terms of a partition function. The particular structure of the partition function $Z_S$ as the ratio of two canonical partition functions, one describing the total system, the other one the bare environment, implies that all  thermodynamic potentials as well as other thermodynamic quantities $X$ following from these potentials by linear operations such as specific heats, susceptibilities, etc., are determined by the difference of the respective quantities of the total system and the bare environment; in short:
\be
X_S = X_{\text{tot}} - X_B\:.
\ee{XXX}

Typical examples for such thermodynamic difference  relations  occur in the determination of solvation or hydration energies, \cite{RouxSimonson} or also in measuring the specific heat of a system strongly interacting  with its environment (e.g. see Refs. \cite{HIT,IHT}), to name but a few. 
The latter examples are taken from quantum mechanics, where
the partition function of the system is determined by the same ratio as in eq. (\ref{ZS}) and consequently the relations (\ref{FZ}) -- (\ref{XXX}) apply \cite{quantumremark}.   

With the equations (\ref{pH*}) and (\ref{ZS}) the Hamiltonian of mean force determines both the statistical and the thermodynamical properties of an open system. The statistical properties, at least in principle, could be obtained from a large set of observations of the considered open system \cite{reconstruction} yielding an estimate for the open system pdf $p_\b(\bx)$. But even if $p_\b(\bx)$ were known exactly it would generally not be possible to  infer the thermodynamics of the open system. This is so  because from $p_\b(\bx)$ only the linear combination of the Hamiltonian of mean force and of the free energy follows in the form
\be
-\b^{-1} \ln p_\b = H^* - F_S\;.
\ee{pHF}
An unambiguous separation of the left hand side into the Hamiltonian of mean force and the free energy can be achieved only if the Hamiltonian of mean force is independent of temperature and, additionally, if the reduced pdf is known for different temperatures.
In the appendix \ref{cc} we demonstrate that the requirement of the above mentioned thermodynamic consistency does not help in finding a unique splitting of the logarithm of the reduced pdf into  the Hamiltonian of mean force and the free energy.

Finally we note that, as a consequence of (\ref{ZS}) and (\ref{US}), the internal energy can be written as
\be
U_S = \langle \frac{\partial}{\partial \b} \b H^* \rangle_S\;,
\ee{HU}
where $\langle \cdot \rangle_S = \int d\G_S \cdot p_\b(\bx) $ is the equilibrium average with respect to the open system. We emphasize that in general the average of $H^*$ fails to fully characterize the internal energy of the system. An additional contribution is present if the Hamiltonian of mean force depends on temperature. As a second note we mention that, even though the average on the right hand side of (\ref{HU}) solely refers to the open system, the Hamiltonian of mean force carries information that, as explained above,  cannot merely be retrieved  from  sole observations of the open system stochastics.

Therefore,  the thermodynamics of a strongly coupled  open system does generally still involve the combined total system-bath system as well as the bare environment.

\section{Fluctuating internal energy} \label{FIE}
The existence of a fluctuating internal energy $E(\bx;\l)$ is a central postulate
of  stochastic energetics \cite{SekimotoPTP,SekimotoLNP} as well as of stochastic thermodynamics \cite{SeifertPRL,SeifertAIP,SeifertRPP}. In these theories  the fluctuating internal energy  is supposed to assign to each point $\bx$ of the system's phase space a local value of the internal energy.
The second argument of $E(\bx;\l)$ denotes one or more  generally time-dependent  parameters $\l$ which only enter the system Hamiltonian and are supposed to be externally controllable.
Moreover, it is assumed that $E(\bx;\l)$ is universal with respect to the statistical properties of the system, i.e., it is independent of the pdf characterizing the system under consideration.
By construction, fluctuations of the deterministic function $E(\bx;\l)$   solely emerge in time as a consequence of the assumed stochastic motion of the state $\bx$. A more general scenario of a  random internal energy field has not been considered in these theories \cite{SekimotoPTP,SekimotoLNP,SeifertPRL,SeifertAIP,SeifertRPP}.

The concept of a fluctuating internal energy recently has been generalized
to the situation of an open system at strong coupling \cite{Seifert}. This generalization is based on two assumptions requiring, (i) that the Hamiltonian of mean force, $H^*(\bx;\l)$, is explicitly known, and, (ii) that the {\it fluctuating} internal energy $E(\bx;\l)$ is functionally related to $H^*(\bx;\l)$ in a similar  
way as its thermodynamic average  
in eq.  (\ref{HU}). In fact, the fluctuating internal energy is defined as
\be
E(\bx;\l) = \frac{\partial}{\partial \b} \big( \b H^*(\bx;\l)\big)\:.
\ee{EH}
The first assumption (i) though runs counter to the general idea of stochastic thermodynamics, postulating that a complete description of all processes related to the open system under consideration can be achieved from an intrinsic point of view in the sense that the functional form of the fluctuating internal energy can be retrieved from observations of the system without further recourse to properties of the environment. In the previous section we demonstrated that a reconstruction of $H^*(\bx;\l)$ on the basis of (experimental) data from the open system is possible in specific cases only.

Yet, in the sequel we shall assume the Hamiltonian of mean force $H^*(\bx;\l)$ to be {\it given} and further exploit this assumption. By construction, the fluctuating internal energy defined by (\ref{EH}) correctly  yields the internal thermodynamic energy as the {\it average} with respect to $p_\b(\bx)$; i.e.,
\be
U_S= \langle E(\bx;\lambda) \rangle_S\:.
\ee{USE}
Note that one would obtain the same result for the equilibrium average from any other function $\tilde{E}(\bx;\l)$, which is defined by
\be
\tilde{E}(\bx;\l) =E(\bx;\l)+ h_E(\bx;\lambda)\:,
\ee{EEf}
where  $h_E(\bx;\l) \in \cN_\b$ with
\be
 \cN_\b \equiv \{h(\bx; \l)| \int d \G_s h(\bx; \l) p_\b (\bx)= 0 \}\:
\ee{Nb}
has a vanishing average in thermal equilibrium.
The requirement that the fluctuating internal energy averaged with respect to the equilibrium pdf $p_\b(\bx)$ yields the
internal energy $U_S$ leaves room for a large ambiguity in defining such a function. In a first step we will concentrate on the choice made in the equation (\ref{EH}). Later on below we shall revert to the objective as to which additional arguments must be invoked  towards the goal to  obtain a uniquely defined fluctuating internal energy. Here we note that  $E(\bx,\l)$ can be given the following alternative form, see the appendix \ref{28} for details:
\be
E(\bx;\l) = \langle H_{\text{tot}}|\bx\rangle - \langle H_B \rangle_B\:,
\ee{EHB}
where $\langle \cdot | \bx \rangle = \int d\G_B \cdot w(\by|\bx) $ denotes a conditional average over the  environmental degrees of freedom. The according conditional pdf $w(\by|\bx)$  is obtained from the total equilibrium pdf $\r_\b(\bx,\by)$ by means of the Bayesian rule, yielding
\be
\begin{split}
w(\by|\bx) &= \frac{\r_\b(\bx,\by)}{\int d\G_B \r_\b(\bx,\by)}\\
&=\frac{e^{-\b\big(H_i(\bx,\by)+H_B(\by)\big)}}{\int d\G_B e^{-\beta\big(H_i(\bx,\by)+H_B(\by)\big)}} \\
&= Z_B^{-1} e^{-\b\big(H_{\text{tot}}(\bx,\by) - H^*(\bx)\big)}\:.
\end{split}
\ee{wyx}

In passing, we mention that $w(\by|\bx)$ describes  a constrained equilibrium of the environment. It specifies  the so termed 'stationary' preparation class \cite{GTH} which is defined as the set of all possible initial preparations of system and environment of the type
\be
\r(\bx,\by) = w(\by|\bx) p^0(\bx)\:,
\ee{rwr}
where $p^0(\bx)$ may be an arbitrary initial pdf of the open system.

The difference between the total energy $H_{\text{tot}}(\bx,\by;\l) = H_S(\bx) + H_i(\bx,\by) +H_B(\by)$ and $E(\bx;\l)$ can then be presented as
\be
\begin{split}
G (\bx,\by) &\equiv H_{\text{tot}}(\bx,\by;\l) - E(\bx;\l)\\
&= \d H_i(\bx,\by) +\d H_B(\bx,\by) +  U_B\:,
\end{split}
\label{G}
\end{equation}
where $U_B=\langle H_B\rangle_B$ denotes the internal energy of the bare bath and $\delta H_i(\bx,\by)$ and $\delta H_B(\bx,\by)$ denote the deviations of the interaction Hamiltonian and the bare bath Hamiltonian, respectively, from their conditional averages, i.e.,
\be
\begin{split}
\d H_i(\bx,\by) &\equiv H_i(\bx,\by) - \langle H_i(\bx,\by)|\bx \rangle \\
\d H_B(\bx,\by) &\equiv H_B(\by) - \langle H_B(\by)|\bx \rangle\:.
\label{dHiB}
\end{split}
\end{equation}
The total averages of these deviations vanish if they are performed with respect to any distribution $w(\by|\bx) p^0(\bx)$ from the stationary preparation class. Note that the system specific part $p^0(\bx)$ may deviate from the equilibrium distribution $p_\b(\bx)$.

Following the spirit of stochastic energetics and stochastic thermodynamics one is led to interpret $G(\bx,\by)$ as the fluctuating energy content of the environment.
However, one also can argue that the contribution from the interaction, $\d H_i(\bx,\by)$, instead of being fully accounted to the environment  should, at least partially be allocated to the fluctuating internal energy. This reasoning then implies that the internal fluctuating energy reads
\begin{equation}
e(\bx,\by;\l) = E(\bx;\l) + \a(\bx) \delta H_i(\bx,\by) +h_E(\bx;\l)\:.
\label{eEa}
\end{equation}
As done with  equation (\ref{EEf}), we allowed for a function $h_E(\bx,\l) \in \cN_\b$.
The weight $\a(\bx)$ determines which fraction of the fluctuating part of the interaction energy counts for the fluctuating internal energy.
Any non-vanishing choice of $\a(\bx)$ renders the fluctuating internal energy a random field due to its resulting dependence on the environmental state $\by$ which is sampled from the conditional pdf $w(\bx|\by)$ of the stationary preparation class.

The average of the fluctuating internal energy $e(\bx,\by;\l)$ with respect to a pdf from the stationary preparation class (spc) turns out to be independent of $\a(\bx)$ and is hence given by
\be
\langle e(\bx,\by;\l \rangle_{\text{spc}} = \langle E(\bx,\l) +h_E(\bx;\l) \rangle_p\:,
\ee{eE}
where $\langle \cdot \rangle_{\text{spc}} = \int d\G_S d\G_B \cdot w(\by|\bx) p(\bx)$ and $\langle \cdot \rangle_p = \int d\G_S \cdot p(\bx)$.
In particular, the thermal equilibrium average of $e(\bx,\by;\l)$ coincides with the internal energy $U_S$.

The most general form of the fluctuating internal energy (\ref{eEa}) leads to a modified energy of the environment following from the requirement that the energy of the total system is the sum of the fluctuating internal energy and the environmental energy. The latter then  becomes
\be
\begin{split}
g(\bx,\by) &= H_{\text{tot}}(\bx,\by) - e(\bx,\by;\l)\\
&=\big(1-\a(\bx)\big) \d H_i(\bx,\by) + \d H_B(\bx,\by)\\
&\quad + U_B - h_E(\bx;\l)\:.
\label{gGa}
\end{split}
\end{equation}
For the average of the environmental energy with respect to a pdf from the stationary preparation class one obtains
\be
\langle g(\bx,\by) \rangle_{\text{spc}} = U_B - \langle h_E(\bx,\l) \rangle_p \;\:.
\ee{avg}
which reduces to $U_B$ in thermal equilibrium.

On the basis of the general forms of the expressions for the fluctuating internal energy and of the energy content of the environment one may identify two additional requirements that lead to a unique form of the internal energy and
consequently also of the environmental energy content.

First, if one demands that the fluctuating internal energy is a function of the system state $\bx$, but is not allowed to depend on the environmental state $\by$ the only choice left  is $\a(\bx) \equiv 0$. Yet, a possible contribution of a function $h_E(\bx) \in \cN_\b$, see eq. (\ref{Nb}), still leaves an ambiguity in the definition of the fluctuating internal energy.

Also $h_E(\bx;\l)$ must vanish if one further requires that the bath energy must not contain  solely $\bx$-dependent terms  other than those entering $G(\bx,\by)$ via the deviations of $H_i(\bx,\by)$ and $H_B(\by)$ from their respective conditional averages. Hence, under these additional assumptions the internal fluctuating energy is  defined by equation (\ref{EH}). Both these requirements possess structural appeal, a deeper physical foundation though is missing.\\

Furthermore, we note that the systems that are typically considered within the frameworks of stochastic energetics and stochastic thermodynamics are characterized by a position-like quantity $\bq$ which undergoes a stochastic motion in a potential energy landscape $V(\bq)$ on the system's configuration space governed by an overdamped Langevin equation, where $\bq$ denotes the configurational component of the phase space point $\bx =(\bq,\bp)$.
The phase space pdf $p_\b(\bx)$ is related to the stationary pdf $p^{\text{conf}}_\b(\bq) = Z^{-1}_{\text{conf}} e^{-\b V(\bq)} $ 
by $p_\b(\bx) = Z^{-1}_{\text{kin}} e^{-\b T(\bp)} p^{\text{conf}}(\bq)$ provided that $H^*(\bx) = T(\bp) + V(\bq)$. This structure follows from a total Hamiltonian of the form $H_{\text{tot}}= T(\bp) +T_B(\by^p) +V_{\text{tot}}(\bq,\by^q)$. 
Here $T(\bp)$ and $T_B(\by^p)$ denote the kinetic energies of the open system and the environment, respectively, while $V_{\text{tot}}(\bq,\by^q)= V_s(\bq) +V_i(\bq,\by^q) +V_B(\by^q)$ is the total potential energy including system, environment and their mutual interaction. 
Further $V(\bq) = -\b^{-1} \ln \i d\by^q e^{-\b V_{\text{tot}}(\bq,\by^q)}/\i d\by^q e^{-\b V_B(\by^q)}$ 
is the so-called potential of mean force; accordingly $Z_{\text{conf}}= \int d\bq e^{-\b V(\bq)}$ and $Z_{\text{kin}}= \int d\bp e^{-\b T(\bp)}$ denote the configurational and kinetic partition functions, respectively. In Ref. \cite {SekimotoLNP} the potential energy entering the Langevin equation is defined as $V^{S}(\bq) = - \b^{-1} \ln \int d\by^q e^{-\b V_{\text{tot}}(\bq,\by^q)}$; it hence differs from the potential of mean force by the configurational part of the partition function of the bare environment. This difference is unimportant for the dynamics because it is  constant with respect to the position $\bq$ and also drops out when energy differences between different configurations at the same temperature are considered. Due to the remaining temperature dependence of the
potential of mean force, however, it must be distinguished from the configurational part of the fluctuating internal energy. In  analogy to (\ref{EH}) this fluctuating internal configuration energy  is given by $E_{\text{conf}}(\bq) = \partial \b V(\bq)/\partial \b$. Upon omitting the $\b \partial V(\bq) /\partial \b$ term one disregards thermodynamic consistency.

\section{Fluctuating nonequilibrium work and heat}\label{Attempt}

As already mentioned, $\l$ is a system's parameter that is supposed to be externally controllable. It enters into the Hamiltonian of the system, $H_S(\bx;\l)$, while the interaction with the environment as well as the Hamiltonian of the bare environment are considered as being  independent of $\l$. When the parameter $\l$ is changed in time, say between the times $t=0$ until $t=\t$,
the nonequilibrium work $w$ applied to the system coincides with
the change of the total Hamiltonian \cite{gauge}, and, hence, is given by
\be
w
=H_{\text{tot}}\big (\bZ(\tau,\bz);\l(\tau)\big ) - H_{\text{tot}}\big (\bz;\l(0) \big)\:,
\ee{w}
provided that the total system initially
stays in the micro-state $\bz=(\bx,\by)$. From there the total system deterministically evolves according to the Hamiltonian equations of motion to the final state $\bZ(\tau,\bz)=\big(\bX(\tau,\bz),\bY(\tau,\bz)\big)$.
We note that the definition of work (\ref{w}) allows for fluctuations which result from the particular choice of the environmental part $\by$ of the initial state $\bz$. Its value is taken from the conditional probability $w(\by|\bx)$ of the stationary preparation class.
Obviously, the work defined by (\ref{w}) cannot be directly inferred from  knowing only the system's states $\bx$ and $\bX(\t,\bx,\by)$ at the beginning and the end of the force protocol. However, when the trajectory $\bX(t,\bx,\by)$
is known for the whole protocol, i.e. for all $t \in [0,\t]$, the work can be obtained as the integral over the supplied power, reading \cite{JarzynskiJSM}
\be
w = \int_0^\t dt \frac{\partial}{\partial \l} H_S\Big(\bX(t,\bx,\by);\l(t)\Big) \dot{\l}(t) \;,
\ee{wit}
where $\dot{\l}$ denotes the time-derivative of $\l$. Here we  used the Hamiltonian relation that
$dH_{\text {tot}}/dt = \partial H_{\text {tot}}/\partial t= \partial H_S/\partial t$.

The definition of  fluctuating heat can be based on the assumption of a particular form for the fluctuating internal energy $e(\bz;\lambda)$ which is supposed to satisfy a first-lawlike balance equation according to which any  change of the internal energy  can be split into the sum of work and heat, reading
\begin{equation}
\D e = w + q\:.
\label{Dewq}
\end{equation}
Because, according to (\ref{w}), the work $w$ is determined by the difference of the total Hamiltonians at the end and the beginning of the force protocol
the fluctuating system heat $q$ can be expressed as the negative difference of the bath energies  defined in (\ref{gGa}). With the particular choice $\a(\bx) \equiv 0$ and $h_E(\bx,\l) \equiv 0$ one has
\begin{equation}
\begin{split}
q(\bz) &= G(\bz) - G\big(\bZ(\tau,\bz)\big)\\
&= -\delta H_i\big(\bZ(\tau,\bz) \big )\\
&\quad - \delta H_B \big(\bZ(\tau,\bz)\big) + \delta H_i(\bz) + \delta H_B(\bz)\:.
\end{split}
\label{qg}
\end{equation}

In particular one assumes here that the form of the fluctuating internal energy remains unchanged even if the system is driven out of equilibrium by the application of a possibly fast and violent action of an external force.  

For a system initially staying in thermal equilibrium, the average heat supplied to the system up to time $t$ becomes
$\langle q \rangle_\b = U_B - \langle G\big(\bZ(t,\bz) \big)\rangle$, where we used that the average initial energy content of the environment is given by $U_B$ because of $\langle G(\bz) \rangle_\b = \langle H_{\text{tot}}(\bz) \rangle_\b - \langle E(\bx;\l)\rangle_\b$ together with $\langle H_{\text{tot}}(\bz) \rangle_\b= U_S +U_B$ and $ \langle E(\bx;\l)\rangle_\b= U_S$. Considering a protocol consisting of a cyclic parameter change of duration $\t_c$ and a subsequent relaxation phase up to the time $\t \gg \t_c$ the system has returned to its initial equilibrium state and hence the environmental energy
becomes $ \langle G\big(\bZ(\t,\bz)\big) \rangle = \langle H_{\text{tot}}\big(\bZ(\t,\bz),\lambda(0)\big) \rangle -U_S$, yielding that $\langle q \rangle_\b = U_{\text{tot}} - \langle H_{\text{tot}}\big(\bZ(\t,\bz),\lambda(0)\big) \rangle = -\langle w \rangle$. This is the expected result: the complete energy supplied to the open system is finally  dumped into the environment. It is interesting to note that this implies that in such a process even the final pdf of the total system must deviate from the form of the initial  stationary preparation class. Otherwise, the resulting heat would vanish on average \cite{heat}. Derivations of second-law like relations which are based on the assumption that, during the relaxation process, the pdf of the total system
stays within the initial stationary preparation class \cite{Seifert} are therefore to be  questioned. \\

In principle, an alternative approach may be based on the control of individual energy fluxes rather than balancing  the total amount of supplied and exchanged
energy. Using the power, the flux corresponding to the supplied work
is given as a function of the system's state alone. This property, however, is 
not shared by the flux of the fluctuating internal energy which is given by the time derivative of the fluctuating energy. As a result it depends on the instantaneous state of the environment. Consequently, also 
the heat-flux defined as the difference between the internal energy flux and the power is explicitly dependent on the state of the environment. Thus, the fluxes of internal energy and of heat are generally inaccessible in an experiment.   
For more details we refer the reader to the appendix \ref{hr}.

\section{Fluctuating entropy and free energy}\label{FEFE}

We next ask whether one may require thermodynamic consistency not only for the averaged quantities, i.e. the  thermodynamic potentials but as well also for their hypothetical fluctuating counterparts  like the fluctuating internal energy $E(\bx;\b)$, fluctuating entropy $s(\bx;\b)$ and the  fluctuating free energy $f(\bx;\b)$. For a reason that will become clear soon we here explicitly display the dependence of these fluctuating quantities on the inverse temperature $\b$ rather than on the parameter $\l$. Stochastic energetics does not consider other fluctuating potentials than internal energy, but the notion of fluctuating entropy is a central element of stochastic thermodynamics. One would assume that then the free energy should also be allowed to fluctuate.

It will turn out that, in general, both the fluctuating free energy and the fluctuating entropy will depend on the state of the reduced system in terms of the momentary pdf $p(\bx)$. The latter may differ from the equilibrium pdf $p_\beta(\bx)$ caused by 
 a nonequilibrium initial state of the open system,  
by a time-dependent forcing $\l(t)$ of the open system or as the   
result of a combination of initial non-stationarity and driving.
In order to be able to assign a Hamiltonian of mean force to the open system the initial state of the environment must be given by the constrained equilibrium $w(\by|\bx)$ characterizing the stationary preparation class. This excludes other initial states taken from different preparation classes \cite{GTH} of the total system, such as  
uncorrelated initial states described by a pdf factorizing into a system and an environmental part, or also systems in contact with heat baths at different temperatures.

Before considering admissible nonequilibrium situations we will first consider the full equilibrium situation in which system and environment stay in their common thermodynamic equilibrium state given by the pdf $\r_\b(\bx,\by)$ defined in equation (\ref{rb}).

\subsection{Fluctuating entropy and free energy: Equilibrium}\label{equi}
Assuming that the hypothetical fluctuating entropy and free energy do not depend on the microscopic state of the environment and hence are functions of the system variable $\bx$ alone we demand
that they yield the corresponding thermodynamic functions upon the average with respect to the equilibrium pdf $p_\beta(x)$ of the open system, i.e.
\begin{align}
\label{SSs}
S_S &= \langle s(\bx;\b) \rangle_S = \int d\G_S s(\bx;\b) p_\b(\bx)\:, \\
F_S   &= \langle f(\bx;\b) \rangle_S = \int d\G_S f(\bx;\b) p_\b(\bx)\:.
\label{FSf}
\end{align}

Thermodynamic consistency requires relations (\ref{UF}) and (\ref{FS})
between $U$ and $F$ as well as between $S$ and $F$, respectively. They can be expressed in terms of the fluctuating quantities, reading:
\begin{flalign}
\label{Ef}
&\int d\G_S p_\b(\bx) \bigg \{ E(\bx;\b) - f(\bx;\b) \nonumber\\
& \bigg . - \b \left [\frac{\partial}{\partial \b} f(\bx;\b) + f(\bx;\b) \frac{\partial}{\partial \b} \ln p_\beta(\bx) \right ] \bigg \}=0 \:,\\
&\int d\G_S p_\b(\bx) \bigg \{ s(\bx;\b) \bigg .  \nonumber  \\
&\bigg . -k_B \b^2\left [ \frac{\partial}{\partial \b} f(\bx;\b) + f(\bx;\b) \frac{\partial}{\partial \b} \ln p_\b(\bx) \right] \bigg \}=0 \:.
\label{sf}
\end{flalign}
Here, the presence of the logarithmic derivatives of the equilibrium pdf $p_\b(\bx)$ resulted when the $\b$-differentiations of the free energy were taken within the average. Consequently we find that
\begin{flalign}
\label{Efx}
& E(\bx;\b) - f(\bx;\b) \nonumber \\
&- \beta \left [\frac{\partial}{\partial \b} f(\bx;\b) + f(\bx;\b) \frac{\partial}{\partial \b} \ln p_\b(\bx) \right ] = h_F(\bx;\b)\:,\\
& s(\bx;\b)-k_B \b^2 \nonumber\\
& \times \left [ \frac{\partial}{\partial \b} f(\bx;\b) + f(\bx;\b) \frac{\partial}{\partial \b} \ln p_\b(\bx) \right] = h_S(\bx;\b)\:,
\label{sfx}
\end{flalign}
where $h_F(\bx;\b)$ and $h_S(\bx;\b)$ are elements from the null-space $\mathcal{N}_\beta$ defined in equation (\ref{Nb}).\\
Equation (\ref{Efx}) equally holds with $\tilde{E}(\bx;\b)$ replacing $E(\bx;\b)$ with an accordingly modified inhomogeneity $h_F(\bx;\b) \in \cN_\b$.\\

It follows from equations (\ref{Efx}) and (\ref{sfx}) in agreement with (\ref{FUTS}) that the fluctuating thermodynamic potentials are related by
 \be
\begin{split}
s(\bx;\b) &= k_B \b \big( E(\bx;\b) - f(\bx;\b) \big)\\
&\quad  +h_S(\bx;\b) -k_B \b h_F(\bx;\b)\:.
\end{split}
\ee{sEf}

As a first example we consider the case $h_S(\bx;\b) = k_B \b h_F(\bx;\b) =0$
for which we find the free energy $f^0(\bx,b)$ as the solution of the differential equation (\ref{Efx}) to become
\be
\begin{split}
f^0(\bx;\b) &= \frac{\b_0 p_{\b_0}(\bx)}{\b p_\b(\bx)} f^0(\bx;\b_0)\\
& \quad +\frac{1}{\b p_\b(\bx)} \int_{\b_0}^\b d\b' p_{\b'}(\bx) \frac{\partial}{\partial \b'} \b' H^*(\bx;\b')
\end{split}
\ee{fp}
where $\b_0$ is an initial inverse temperature and $f^0(\bx,\b_0)$ a reference fluctuating free energy at this inverse temperature. The equilibrium average of this reference free energy must agree with the thermodynamic free energy $F_S$ at the inverse temperature $\b_0$ but otherwise $f^0(\bx,\b_0)$ is arbitrary. The corresponding fluctuating entropy $s^0(\bx,\b)$ follows from Eq. (\ref{sEf}) as
 \begin{equation}
s^0(\bx;\b) = k_B \b \big ( E(\bx;\b) - f^0(\bx;\b) \big )\:.
\label{sEf0}
\end{equation}
We note that this form of fluctuating entropy differs from the fluctuating entropy $s^{\text{sth}}(\bx;\b)$ introduced in Ref. \cite{Seifert}; the latter reads instead
\be
s^{\text{sth}}(\bx;\b) = s_0(\b) - k_B \ln p_\b(\bx) + k_B \b^2 \frac{\partial}{\partial \b} H^*(\bx;\b)\:,
\ee{ssth}
where $s_0(\b)$ is an unspecified constant that may only depend on the inverse temperature but not on the phase space variable $\bx$ \cite{diff}.
Using equation (\ref{pHF}) we express $\ln p_\b(\bx)$ in terms of $H^*(\bx;\b)$ and $F_S$, yielding
\be
s^{\text{sth}}(\bx;\b) = k_B\b \big( E(\bx;\b) - F_S(\b) \big) + s_0(\b) \:.
\ee{sEF}
Now, as a necessary condition for thermodynamic consistency the entropy constant $s_0(\b)$ must vanish.
Furthermore, a comparison of the two fluctuating entropy expressions (\ref{sEf}) and (\ref{sEF}) implies that
 the fluctuating free energy $f^{\text{sth}}(\bx;\b)$ must coincide with the ({\it non-fluctuating}) thermodynamic free energy $F_S(\b)$, i.e.
\be
f^{\text{sth}}(\bx;\b) = F_S(\b)\:.
\ee{fF}
Yet, the choice (\ref{EH}), (\ref{ssth}) and (\ref{fF}) is thermodynamically consistent if one chooses
\be
\begin{split}
h_F(\bx;\b) &= E(\bx;\b) -\frac{\partial}{\partial \b} \b F_S(\b)\\
&\quad + \b F_S(\b) \frac{\partial}{\partial \b} \ln p_\b(\bx)
\end{split}
\ee{hF}
and
\be
h_S(\bx;\b) = - k_B \b h_F(\bx;\b)\:.
\ee{hS}
Using (\ref{UF}), (\ref{USE}) and $\langle \partial \ln p_\b(\bx) /\partial \b \rangle_S =0$, we confirm that the average of $h_F(\bx;\b)$ with respect to equilibrium pdf $p_\b(\bx)$ vanishes and hence $h_F(\bx;\b) \in \cN_\b$.

The choices $h_F(\bx;\b)= h_S(\bx;\b)=0$ and (\ref{hF}, \ref{hS})) are just two examples from a continuum of thermodynamically consistent families of fluctuating internal and free energies and fluctuating entropy labeled by functions $h_E(\bx;\b), h_F(\bx;\b), h_S(\bx;\b)$, each of which must be an element of $\cN_\b$, but otherwise can be arbitrary.

The fact that the particular form (\ref{ssth}) of the fluctuating entropy in \cite{Seifert} implies, via thermodynamic consistency, a fluctuating free energy that coincides with the thermodynamic free energy might be considered as a curiosity, which went unnoticed in \cite{Seifert}.  

\subsection{Fluctuating entropy and free energy: Nonequilibrium}\label{nequi}
The kind of nonequilibrium situations considered here is restricted to systems interacting with a single heat bath. The total system must initially stay within the stationary preparation class specified by the conditional environmental pdf $w(\by|\bx)$ given by equation (\ref{wyx}). As indicated above, a nonequilibrium situation may result from a non-stationary initial state or a time-dependent variation of the external control parameter $\l$. Within the  equilibrium preparation class, the system may initially assume any pdf
$p^0(\bx)$ leading to the initial pdf at time $t=0$,
$\r^{\text{ini}}(\bx,\by) = w(\by|\bx) p^0(\bx)$, see (\ref{rwr}).
Starting from there it evolves according to the Hamiltonian dynamics of the total system and, at a later time $t$, yields the reduced pdf $p^t(\bx)$ given by
\be
p^t(\bx) = \int d \G_B p^0 \big(\bX(-t,\bz)\big) w\big(\bY(-t,\bz)|\bX(-t,\bz)\big)\:,
\ee{pt}
where as in equation (\ref{w}) $\bZ(t,\bz) =\big(\bX(t,\bz),\bY(t,\bz)\big)$ denotes the time evolution in the phase space of the total system starting at $\bz=(\bx,\by)$ under the Hamiltonian dynamics of the total system. The fact that $w(\by|\bx)$ depends on temperature renders also $p^t(\bx)$ temperature dependent for all times $t>0$, even if $p^0(\bx)$ was independent of temperature.
In passing we note that the joint pdf $\r^t(\bx,\by)$ which is given by the integrand of (\ref{pt}) in general does not have the form $p^t(\bx) w(\by|\bx)$ for times $t>0$.
 
Because in the sequel we will consider the open system only at a fixed time $t$ we suppress the time-dependence of the system's pdf for the sake of simplifying our notation, and instead emphasize the $\b$-dependence in writing
\be
 p(\bx;\b) \equiv p^t(\bx)\:.
\ee{ppt}

The extension of the structure of thermodynamics as it applies to equilibrium systems to nonequilibrium situations requires to postulate the existence of nonequilibrium thermodynamic potentials $U^\neqq(\b)$, $F^\neqq(\b)$ and $S^\neqq(\b)$ satisfying the thermodynamic relations (\ref{US}) -- (\ref{FUTS}), guaranteeing a corresponding thermodynamic consistency. In stochastic thermodynamics the nonequilibrium internal energy is determined by the average of the same fluctuating internal energy as in thermal equilibrium; however, the average being taken now with respect to the actual reduced pdf $p(\bx;\b)$ given by (\ref{ppt}). Hence
\be
U^\neqq_S = \int d\G_S E(\bx;\b) p(\bx;\b) \;,
\ee{UEp}
with $E(\bx;\b) = \partial \beta H^*(\bx;\b) /\partial \b$.
Expressing also $F^\neqq(\b)$ and $S^\neqq(\b)$ as averages of yet undetermined respective fluctuating quantities $f^\neqq(\bx;\b)$ and $s^\neqq(\bx;\b)$ taken with respect to $p(\bx;\b)$ we deduce from the requirement of thermodynamic consistency
the following two equations
\begin{flalign}
\label{Efxn}
&E(\bx;\b) - f^\neqq(\bx;\b) \\
&- \beta \left [\frac{\partial}{\partial \b} f^\neqq(\bx;\b) + f^\neqq(\bx;\b) \frac{\partial}{\partial \b} \ln p(\bx;\b) \right ] = h_F(\bx;\b)\:,\nonumber\\
&s^\neqq(\bx;\b) -k_B \b^2 \label{sfxn}\\
& \times \left [ \frac{\partial}{\partial \b} f^\neqq(\bx;\b) + f^\neqq(\bx;\b) \frac{\partial}{\partial \b} \ln p(\bx;\b) \right] = h_S(\bx;\b) \:,\nonumber
\end{flalign}
where the averages of the functions $h_F(\bx;\b)$ and $h_S(\bx;\b)$ with respect to $p(\bx;\b)$ vanish, otherwise these functions may be chosen arbitrarily, i.e. $h_F(\bx;\b), h_S(\bx;\b) \in \cN^\neqq$ with $\cN^\neqq = \{h(\bx)| \int d\G_S h(\bx) p(\bx;\b) =0 \}$.

For any particular choice of $h_F(\bx;\b)$ and $h_S(\bx;\b)$ the first equation determines a fluctuating free energy and the second one a fluctuating entropy in analogy to the equilibrium situation described by the equations (\ref{Efx}) and (\ref{sfx}).
Accordingly, also equation (\ref{sEf}) holds for the nonequilibrium potentials:
 \be
\begin{split}
s^\neqq(\bx;\b)& = k_B \b \big ( E(\bx;\b) - f^\neqq(\bx;\b) \big )\\
& \quad +h_S(\bx;\b) -k_B \b h_E(\bx;\b)\:.
\end{split}
\ee{sEfn}
Choosing $h_S(\bx;\b) = k_B \b h_F(\bx;\b)=0$ we find for the fluctuating free energy and entropy the same expressions (\ref{fp}) and (\ref{sEf0}), respectively, with the reduced equilibrium pdf $p_\b(\bx;\b)$ replaced by the actual pdf $p(\bx;\b)$. In general, one may obtain a thermodynamically consistent theory by prescribing any form of either the free fluctuating energy or of the fluctuating entropy. The other fluctuating potential together with the auxiliary functions can then be determined by any two of the equations (\ref{Efxn}),  (\ref{sfxn})  and (\ref{sEf}). For the technical details see  the appendix \ref{construction}.

\section{Summary and Conclusions} \label{Con}
We investigated the problem whether a classical open system may be described in terms of {\it fluctuating} thermodynamic potentials. Provided that the open system is in equilibrium with its environment at an inverse temperature $\b$ these fluctuating thermodynamic potentials are required to yield the standard thermodynamic potential on average. We thus first recapitulated the thermodynamics and statistical mechanics of systems interacting with an environment at any coupling strength. In this context, of central importance is an effective system Hamiltonian which is renormalized by the presence of the environment and which is referred to as Hamiltonian of mean force, $H^*$. It determines the equilibrium distribution of the system in its phase space being proportional to $e^{-\b H^*}$ with a normalizing proportionality factor yielding the  partition function of the open system.

In general, the average over the environmental degrees of freedom entails a temperature dependence of the Hamiltonian of mean force. This temperature dependence makes it impossible to infer the Hamiltonian of mean force from the sole knowledge of the normalized equilibrium distribution, which in principle is accessible from measuring the stochastic system trajectories experimentally. Put differently, the Hamiltonian of mean force cannot be determined by solely monitoring the open system dynamics. 

The internal energy of the open system is related to the equilibrium average of the Hamiltonian of mean force and its $\b$-derivative, see equation  (\ref{HU}). Starting from this expression one may hypothesize the form of the corresponding fluctuating internal energy, which in general contains a large amount of arbitrariness. One obtains the defined functional form of the fluctuating internal energy in agreement with Ref. \cite{Seifert} if one requires (i) that this fluctuating internal energy only depends on system variables but not on the  environmental degrees of freedom, 
and (ii) that the energy assigned to the environment must not contain an additive contribution that only depends on system variables other than deviations of the interaction Hamiltonian and environmental Hamiltonian from their conditional expectation values. The first requirement implies that the weight $\a(\bx)$ in (\ref{gGa}) vanishes and the second one that $h_E(\bx,\l) =0$. The resulting form is denoted by $E(\bx,\l)$ and given by equation (\ref{EH}). One must keep in mind however that it cannot be determined from a purely system intrinsic point of view as it carries information that does not follow from the distribution of system phase space points in equilibrium, see the Appendix \ref{hr} and in particular equation (\ref{dq2}).

The work that is supplied to the system by means of a variation of externally controlled parameters is well defined. It is a fluctuating quantity that can be determined on the basis of a stochastic trajectory of the system extending over the duration of the force protocol. 
If one adopts $E(\bx,\l)$ as the definition of a fluctuating internal energy and postulates  further a first-lawlike relation also a fluctuating heat can be assigned to stochastic trajectories of the open system. In this approach work and heat refer to the finite changes of the respective energies over some protocol of finite duration. An alternative approach based on the balance of instant individual energy fluxes is not feasible because the fluxes of the fluctuating internal energy and of the heat explicitly depend on the environmental degrees of freedom and hence are inaccessible from a system intrinsic point of view. Hence, the open-system-intrinsic point of view proves to be insufficient to identify internal energy and heat independent of whether finite changes or fluxes of these quantities are considered.

The example of a protocol consisting of a short cyclic parameter variation followed by a long relaxation period demonstrates that even at large times the environment does not return to the initial state described by the stationary preparation class.
It is therefore quite questionable to assume that the pdf of the total system driven by a parameter change will permanently stay in the stationary preparation class. The microscopic analysis of the total system leads one to the same conclusion.
The derivation of a second-lawlike relation in Ref. \cite{Seifert} is however based on this doubtful assumption.

We further studied whether the requirement of thermodynamic consistency allows one to specify fluctuating thermodynamic potentials other than a fluctuating internal energy such as a fluctuating free energy and a fluctuating  entropy. The latter plays a central role in stochastic thermodynamics. It turns out that thermodynamic consistency is not particularly restrictive and leaves the possibility to prescribe the fluctuating entropy or the fluctuating free energy in a virtually arbitrary way.

In summary, we note that the specific relations between thermodynamic potentials can be extended to corresponding fluctuating quantities and, in this way, a consistent stochastic thermodynamics can be constructed for open systems  in contact with a single heat bath at a prescribed temperature. 
This framework comprises nonequilibrium situations caused by initial system states differing from equilibrium or by external, time-dependent forcing.

For other nonequilibrium situations \cite{GTH,GHT80} which, for example, are caused by initially uncorrelated states of the system and the environment,  or by the coupling to two heat-baths at different temperatures a Hamiltonian of mean force cannot be defined. Therefore, these situations cannot be described within the framework of stochastic thermodynamics \cite{remark}.

Yet in those situations that can be characterized by the stationary preparation class, the lack of further physical principles leaves one with an enormous ambiguity. The virtual arbitrariness in defining fluctuating entropies or fluctuating free energies raises serious  doubts concerning the physical relevance of such theories.

\acknowledgments
PT thanks the
Foundation for Polish Science (FNP) for providing him with an Alexander von Humboldt Polish Honorary
Research Fellowship.

\appendix
\section{Hamiltonian of mean force and thermodynamic consistency}\label{cc}

We demonstrate that thermodynamic consistency does not suffice to construct the Hamiltonian of mean force $H^*(\bx)$ and the corresponding free energy $F_S$ from a given equilibrium pdf (\ref{pH*}) which may be written as $p_\b(\bx) = e^{-\b H^*_0(\bx)}$.
This holds because any separation $H^*_0(\bx) = H^*_x(\bx) -F_x$ implies a partition function  $Z_x = \int d\G_S e^{-\b H^*_x(\bx)}=e^{-\b F_x}$, which entails thermodynamically consistent  potentials $F_x$, $U_x$ and $S_x$.
Also the equation (\ref{HU}), which  connects the Hamiltonian of mean force and thermodynamics, does not provide extra information about a proper separation of $H^*_0(\bx)$.

To understand this we substitute $H^*(\bx)= H^*_0(\bx) +F_S$  in (\ref{HU}) and obtain
\be
\begin{split}
U_S &= \langle \frac{\partial}{\partial \b} \b H^*_0 \rangle_S +\frac{\partial}{\partial \b} (\b F_S)\\
&= U_S\:.
\end{split}
\ee{USUS}
The second line follows because of $\partial( \b F_S)/\partial \b =U_S$, see  (\ref{UF}), and  $\langle \partial \b H^*_0 /\partial \b \rangle_S = -\frac{\partial}{\partial \beta}  \int d \G_S e^{-\b H^*_0(\bx)}  = -\frac{\partial}{\partial \beta} 1 = 0$. In conclusion, we find that (\ref{HU}) reduces to an identity and hence does not provide a unique identification of $H^*$.

Put differently, thermodynamic consistency does not impose a condition on a proper separation of $H^*_0$. For any given free energy $F_S$ the relations (\ref{UF}) and (\ref{FS}) define the internal energy and the entropy, respectively. As a consequence equation (\ref{FUTS}) follows.

\section{Derivation of equation (\ref{EHB})}
\label{28}
Starting from the definition (\ref{EH}) of $E(\bx;\lambda)$, and combining with the expression (\ref{H*}) we obtain
\be
E(\bx) = H_S(\bx) - \frac{\partial}{\partial \b} \ln \langle e^{-\b H_i(\bx,\by)} \rangle_B\:.
\ee{EHH}
The second part on the right hand side can be further evaluated to yield
\be
\begin{split}
\frac{\partial}{\partial \b} \ln \langle e^{-\b H_i} \rangle_B &= \frac{ \partial \langle e^{-\b H_i} \rangle_B / \partial \b}{\langle e^{-\b H_i} \rangle_B}\\
&= -\frac{\int d\G_B \big (H_i(\bx,\by) +H_b(\by) \big ) e^{-\b (H_i+H_B)}}{\int d\G_B e^{-\beta (H_i+ H_B)} }\\
& \quad
+\frac{\int d\G_B H_B(\by) e^{-\b H_B(\by)}}{\int d\G_B e^{-\b H_B(\by)}}\\
&= -\langle \big ( H_i(\bx,\by) +H_B(\by) \big ) |x \rangle +U_B\:,
\label{lneHi}
\end{split}
\end{equation}
where the conditional average $\langle \cdot |\bx \rangle$ is performed with respect to the conditional probability $w(\by|\bx)$ defined in (\ref{wyx}).
Combining (\ref{EHH}) and (\ref{lneHi}) one obtains (\ref{EHB}).

\section{Heat rate} \label{hr}
For the sake of simplicity we here use $E(\bx,\l)=\partial \big( \b H^*(\bx;\l)\big)/ \partial \b$  as an expression for the fluctuating internal energy. We further assume that $\l$ is a function of time  and consider the time rate of change of $E(\bx;\l)$. According to the underlying Hamiltonian motion of the total system $d E(\bx;\l)/dt$
is given by the Hamiltonian equations of motion reading in this case
\be
\frac{d}{dt} E(\bx;\l) = \{H_{\text{tot}}(\bx,\by), E(\bx; \l) \} + \frac{\partial}{\partial \l} E(\bx;\l) \:\dot{\l}\:,
\ee{dEdt}
where
\be
\{f(\bx,\by),g(\bx)\} = \sum_i \frac{\partial f}{\partial x^p_{i}}\frac{\partial g}{\partial x^q_{i}} - \frac{\partial f}{\partial x^q_{i}}\frac{\partial g}{\partial x^p_{i}}
\ee{Pb}
denotes the Poisson bracket of two phase space function $f(\bx,\by)$ and $g(\bx)$;
further, $x^q_i$ and $x^p_i$ denote the position and momentum components, respectively, of the $i$-th degree of freedom of the open system. Derivatives with respect to the environmental positions and momenta are absent because the fluctuating internal energy $E(\bx;\l)$ only depends on $\bx$ by assumption.
In the second term on the right hand side of (\ref{dEdt}) $\dot{\lambda}$ denotes the time derivative of the parameter $\l$.
This second term may be transformed as follows:
\be
\begin{split}
\frac{\partial}{\partial \l} E(\bx;\l) &= \frac{\partial}{\partial \l} \frac{\partial}{\partial \b} \b H^*(\bx;\l)\\
&=\frac{\partial}{\partial \b} \b \frac{\partial H^*(\bx;\l)}{\partial \l} \\
&=\frac{\partial H_S(\bx;\l)}{\partial \l}\:.
\label{El}
\end{split}
\end{equation}
Hence, the second term
on the right hand side of equation (\ref{dEdt}) agrees with the time rate of change of the work done on the open system by the changing external parameter $\l$ and hence coincides with the power supplied to the system, which is given by
\be
\dot{w} = \frac{\partial H_S(\bx;\l)}{\partial \l} \dot{\l}\:.
\ee{dw}
According to the first-lawlike equation (\ref{Dewq}), now written in the time local form $\dot{e} = \dot{w} +\dot{q}$, the first term on the right hand side of equation (\ref{dEdt}) gives the time rate of change of the heat $\dot{q}$ that is exchanged between system and environment.
This  heat flux hence becomes
\be
\dot{q}(\bx,\by) = \{ H_{\text{tot}}(\bx,\by;\l), \langle H_{\text{tot}}(\bx,\by)|\bx \rangle \}\:,
\ee{dqHH}
where we expressed the fluctuating internal energy with the help of  equation (\ref{EHB}) in terms of the conditional average of $H_{\text{tot}}(\bx,\by)$.
Using equation (\ref{wyx}) one may write the heat flux as
\be
\begin{split}
\dot{q}(\bx,\by) &= \{H_{\text{tot}}(\bx,\by;\l),H^*(\bx;\l) \} \b \langle H_{\text{tot}}(\bx,\by;\l)|\bx \rangle\\
&\quad+\int d\G'_B \{ H_{\text{tot}}(\bx,\by;\l), H_{\text{tot}}(\bx,\by';\l)\}\\
&\quad \times \big [1-\b H_{\text{tot}}(\bx,\by';\l) \big] w(\by'|\bx)\:.\\
\end{split}
\ee{dq2}
Here, the $d \G'_B$ integration refers to the environmental $\by'$ phase space variables.

In contrast to the power which, apart from its $\l$-dependence, is only a function of the system variable $\bx$, the heat flow also depends on the  microscopic state $\by$ of the environment and therefore is an explicitly stochastic object, not only due to the randomness of the system trajectory.
Therefore the heat flux cannot be inferred upon exclusively observing  the stochastic open system dynamics.

\section{Construction of families of thermodynamically consistent fluctuating potentials}\label{construction}
The fluctuating potentials $E(\bx;\b)$, $s^\neqq(\bx;\b)$ as well as $f^\neqq(\bx;\b)$ are thermodynamically consistent if functions $h_F(\bx;\b),\;h_S(\bx;\b) \in \cN_p$ exist such that the equations (\ref{Efxn}) and (\ref{sEf}) are fulfilled. We assume that the internal energy $E(\bx;\b)$ is given and  moreover that the fluctuating entropy is specified up to an additive contribution $c(\b)$. The latter is independent of the phase space coordinate $\bx$ and hence does not fluctuate. In other words, we assume that the functional form of  fluctuating entropy difference between two phase space points is known. Hence, we can write
\be
s^\neqq(\bx;\b) = s^\neqq_0(\bx;\b) + c(\b)\:.
\ee{ssc}
The null-functions $h_F(\bx,\b)$ and $h_S(\bx,\b)$ can formally be absorbed into redefined fluctuating internal energy and entropy, defined as
\begin{align}
\label{tildes}
\tilde{s}^\neqq(\bx;\b)& = s^\neqq(\bx;\b) - h_S(\bx;\b)\:,\\
\tilde{E}(\bx;\b)& =E(\bx;\b) - h_F(\bx;\b)
\label{tildeE}
\end{align}
and consequently we can write (\ref{Efxn}) and (\ref{sEf}) as
\begin{align}
\label{tildeEf}
\tilde{E}(\bx;\b)& = \frac{\partial}{\partial \b} \b f^\neqq(\x;\b) + \b f^\neqq(\bx;\b) \frac{\partial}{\partial \b} \ln p(\bx;\b)\:,\\
\tilde{s}^\neqq(\bx;\b) &= k_B \b \big (\tilde{E}(\bx;\b) - f^\neqq(\bx;\b) \big )\:.
\label{tildesE}
\end{align}
Expressing the fluctuating free energy with the help of (\ref{tildesE}) in terms of  $\tilde{E}(\bx;\b)$ and $\tilde{s}^\neqq(\bx;\b)$ and putting the result in equation (\ref{tildeEf}) we obtain after some algebra
\be
k_B \b \frac{\partial}{\partial \b} p(\bx;\b) \tilde{E}(\bx;\b) = \frac{\partial}{\partial \b} p(\bx;\b) \tilde{s}^\neqq(\bx;\b)\:.
\ee{pEps}
Integrating over the system's phase space $\G_S$ we may express the yet unknown non-fluctuating entropy constant $c(\b)$ in terms of the  averages $U^\neqq(\b) = \int d\G_s p(\bx;\b) E(\bx;\b)$ and $S^\neqq_0(\b) = \int d\G_S p(\bx;\b) s^\neqq_0(\bx;\b)$ reading
\be
c_0(\b) = c_0 - S^\neqq_0 + k_B \int_{\b_0}^\b d \b' \b' \frac{\partial}{\partial \b'} U^\neqq(\b')\:.
\ee{c0}
Because $E(\bx;\b)$ and $s^\neqq_0(\bx;\b)$ are assumed to be known their averages with respect to the also  supposed to be given nonequilibrium pdf $p(\bx;\b)$ can be calculated. We used that their averages coincide by definition with those of the respective auxiliary quantities carrying a tilde.  Eq. (\ref{pEps}) is tantamount to
\begin{flalign}
&k_B \beta \frac{\partial}{\partial \b} p(\bx;\b) E(\bx;\b) - \frac{\partial}{\partial \b} p(\bx;\b) s^\neqq(\bx;\b) = \nonumber\\
& -k_B \beta \frac{\partial}{\partial \b} p(\bx;\b) h_F(\bx;\b) + \frac{\partial}{\partial \b} p(\bx;\b) h_S(\bx;\b)\:. \label{hEhS}
\end{flalign}
We are now free to choose
\be
h_S(\bx;\b) = k_B \b h_F(\bx;\b)\:.
\ee{hShE}
The right hand side of (\ref{hEhS}) then simplifies to yield
\be
\b \frac{\partial}{\partial \b} p(\bx;\b) E(\bx;\b) - k^{-1}_B \frac{\partial}{\partial \b} s^\neqq(\bx;\b) = p(\bx;\b) h_F(\bx;\b)\:.
\ee{hE}
This determines $h_F(\bx;\b)$ in line with the requirement $h_F \in \cN^\neqq$.
Hence, any pair of fluctuating internal energy and fluctuating entropy 
can be complemented by a fluctuating free energy in a thermodynamically consistent way.

One may also prescribe the fluctuating internal energy together with a fluctuating free energy $f^\neqq(\bx;\b) = f^\neqq_0(\bx;\b) - g(\b)$, which still contains a yet unspecified additive contribution $g(\b)$ that is independent of the phase space variable $\bx$. The form of this additive contribution can be found from the requirement that the nonequilibrium averages of $E(\bx;\b)$ and $f^\neqq(\bx;\b)$ satisfy the relation (\ref{UF}). Putting the resulting known fluctuating free energy into (\ref{Efxn}) one obtains the adequate function $h_F(\bx;\b)$. Finally, the relation (\ref{sEf}) yields a whole family of fluctuating entropies whose members differ in the choice of $h_S(\bx;\b)$. By construction, the obtained fluctuating potentials are thermodynamically consistent.


\begin{thebibliography}{99}
\bibitem{quantumremark} Also the other relations of Section \ref{OS} remain valid upon proper replacements of Hamiltonian functions by operators, pdfs by density matrices and phase space integrals by traces $\Tr_S$, $\Tr_B$ over the respective system and environment Hilbert spaces. In the defining equation (\ref{eH*}) of the Hamiltonian of mean force the environmental average on the right hand side has to be understood as $\langle e^{-\b(H_S+H_i)}\rangle_B = Z_B^{-1} \Tr_B e^{-\b(H_S+H_i+H_B)}$, where $Z_B=\Tr_B e^{-\b H_B}$. Finally, equation (\ref{H*}) becomes $H^* = - \b^{-1} \ln (\Tr_B e^{-\b(H_S+H_i+H_B)}/Z_B^{-1})$.
\bibitem{reconstruction} A reconstruction of $p_\b(\bx)$ would require to repeatedly determine the state $\bx \in \G_S$ of the system in equilibrium and to build a histogram from these data as an estimator of the pdf $p_\b(\bx)$. It is obvious that such a procedure becomes  a prohibitively demanding task even for systems with relatively few degrees of freedom.
\bibitem{gauge} For non-autonomous Hamiltonian systems not any Hamiltonian rendering the correct equations of motion gives a proper expression for the energy. However, with a convenient gauge transformation it can always be achived that the value of the Hamiltonian corresponds to the instantaneous energy of the system \cite{CHT}.
\bibitem{heat}Note that the initial state of system plus environment would exactly be approached only if a ``super-bath'' were present. In this case the ''missing'' energy would finally be deposited  there. Even though the difference of the asymptotic states in presence and in absence of a super-bath may be arbitrarily small their energy content differs by a finite amount being relevant for the energetics of the open system.
\bibitem{diff} In Ref. \cite{Seifert} only the difference of fluctuating entropies at distinct phase space points at the same temperature is defined.
\bibitem{remark}
Even if starting with the equilibrium pdf $p_{\b}(\bx)$  in the correlation-free  preparation class this very equilibrium pdf will not stay stationary, but is undergoing  an open nonequilibrium relaxation dynamics which, assuming ergodicity, will approach the equilibrium dynamics in the asymptotic long time limit only. Before equilibrium is established 
one will observe non-stationary correlations between the system stochastics and the  stochastic bath dynamics, which are different from the system-bath stationary equilibrium correlations.
\end{thebibliography}
\end{document}